\documentclass{article}
\usepackage{spconf,amsmath,graphicx}
\usepackage{booktabs}
\usepackage{multirow}
\usepackage[export]{adjustbox}

\title{GENERATING ADVERSARIAL SAMPLES FOR TRAINING WAKE-UP WORD DETECTION SYSTEMS AGAINST CONFUSING WORDS}
\name{Haoxu Wang$^{1,2}$\sthanks{The first two authors contributed equally to this work.}, Yan Jia$^{2*}$, Zeqing Zhao$^{3}$, Xuyang Wang$^{3}$, Junjie Wang$^{3}$, Ming Li$^{1,2}$\sthanks{Corresponding Author.}}
\address{$^{1}$School of Computer Science, Wuhan University, Wuhan, China \\
$^{2}$Data Science Research Center, Duke Kunshan University, Kunshan, China \\
$^{3}$AI Lab of Lenovo Research, Beijing, China \\
ming.li369@duke.edu}
\begin{document}
\ninept
\maketitle
\begin{abstract}
Wake-up word detection models are widely used in real life, but suffer from severe performance degradation when encountering adversarial samples. In this paper we discuss the concept of confusing words in adversarial samples. Confusing words are commonly encountered, which are various kinds of words that sound similar to the predefined keywords. To enhance the wake word detection system's robustness against confusing words, we propose several methods to generate the adversarial confusing samples for simulating real confusing words scenarios in which we usually do not have any real confusing samples in the training set. The generated samples include concatenated audio, synthesized data, and partially masked keywords. Moreover, we use a domain embedding concatenated system to improve the performance. Experimental results show that the adversarial samples generated in our approach help improve the system's robustness in both the common scenario and the confusing words scenario. In addition, we release the confusing words testing database called HI-MIA-CW for future research.
\end{abstract}
\section{Introduction}
\label{sec:intro}

In intelligent speech processing applications, the Keyword Spotting (KWS) system, including wake-up word detection, plays an important role in human-computer interaction. KWS aims to detect a predefined keyword or a set of keywords in a continuous audio stream. 
Studies have been proposed to deliver robust approaches with high detection accuracy, 
etc. Alan and Robert adopted dynamic time warping (DTW) for keyword spotting back in 1985 \cite{1}, then hidden Markov models (HMM)\cite{2,3,4}, deep neural networks (DNN)\cite{Sun2017CompressedTD, Panchapagesan2016MultiTaskLA,5} and other various neural network structures including convolutional neural networks (CNN) \cite{Sainath2015ConvolutionalNN}, temporal convolutional neural networks \cite{TemporalCnn1,TemporalCnn2}, recurrent (RNN) neural networks\cite{10.1007/978-3-540-74695-9_23,WOLLMER2013252} and Transformer\cite{kwsTransformer} are also proposed for this task.
However, 
the probability of false alarm becomes higher under complex acoustic environments and ambiguous content. Without further adaptation, the KWS system may misclassify filler as keywords since some of the filler actually sound close to the keywords.  Those are the adversarial samples which are called confusing words (CW). 
Moreover, it is expensive to acquire human recorded adversarial samples for training a KWS system that can accurately classifies confusing words, especially when the keywords are customized by the users.

In this paper, we discussed the concept of confusing words. And we will release a supplemental database called HI-MIA-CW which we used the same setup of the HI-MIA\cite{9054423} database to record.
There are about 16k audios with 12 confusion words patterns in table \ref{confusion_word} for the keyword in HI-MIA database from new 30 speakers. This data is included in our evaluation set. Then we propose several methods to generate some adversarial samples for simulating real confusing words employed on an end-to-end approach to address the aforementioned issue. The idea is motivated by the maximum mutual information (MMI) criterion 
to improve the discriminative power of the model\cite{Povey+2016}. The first technique is to concatenate the waveform of the real subword audio. 
The second adversarial samples augmentation is performed by a text-to-speech (TTS) system. The use of synthesized speech for data augmentation is also not new\cite{rygaard2015using,mimura2018leveraging,TurkishSyn}. 
\cite{lin2020training,huang2021synth2aug} shows that synthetic data can help train the keyword spotting model. 
The third method is that apply random masking on speech signals to simulate confsuing words like keyword audio that are interrupted by mute in the middle. Moreover, we use domain embeddings extracted from pre-trained LSTM domain classifier to help overcome the domain shift problems. To the best of our knowledge, we are the first one to discuss the concept of confusing words in KWS scenarios and explore the augmentation methods to generate adversarial samples of confusing words to improve the performance of the wake-up word detection system. Both augmentation methods achieve significant improvement on the end-to-end KWS model. Especially for TTS augmentation, the false rejects rate drops from 68.60\% to 9.81\% 
at twenty false alarms in one hours compared with the one that is evaluated with the system trained without the aforementioned confusing words  augmentation approaches. 

This rest of the paper is organized as follows. Section 2 discuss the confusing words and release the database. Section 3 describes the framework of the CNN based KWS system. Section 4 presents our augmentation methods. Section 5 discusses the experimental results, and the conclusion is provided in section 6.

\section{Confusion words}
The definition of confusion words changes with the application scenario. In the natural language processing (NLP) scenarios, some confusing words with similar meanings, but different spellings and pronunciations, will appear in similar contexts and some other confusing words are misspelled. 
\cite{wang2013automatic} present a system named Automatic Confusion words Extraction (ACE), which takes a Chinese word as input and automatically outputs its easily misspelled confused words. 

Unlike confusion words in the NLP domain, confusion words in the speech domain are those words that sound similar to predefined keywords or is the part of the keywords. Let's take the keyword "ni hao, mi ya"(Hello Mia) as an example. Based on the idea of similar pronunciation and fragmentation, we came up with the following confusing words in table \ref{confusion_word}. Table \ref{confusion_word} show the phoneme sequence of the keyword and confusion words, where the subscript of phoneme represents tone. Confusion words as adversarial samples attack the acoustic model in the wake-up word system, cause false rejections that severely disrupt the usage experience. 

In order to compare the performance of models in practical applications, we used the same setup of HI-MIA database \cite {9054423} to further record 16k audios with the same 12 confusion word patterns in table \ref{confusion_word} from new 30 speakers. The supplemental database HI-MIA-CW is released.\footnote{https://github.com/Mashiro009/HI-MIA-CW}\footnote{http://openslr.org}

\begin{table}[h]
    \small
    \setlength{\tabcolsep}{1.0mm}{
    \centering
    \caption{Phoneme sequences of the keyword and confusion words}        
    \label{confusion_word}
    \begin{tabular}{ccc}
        \toprule
        Data Type & Words & Phoneme Sequence   \\
        \midrule
        Keyword & ni hao, mi ya  & N $\text{I}_2$ H $\text{A}_3$ $\text{U}_3$ M $\text{I}_3$ YH $\text{I}_4$ $\text{A}_4$   \\
        \midrule
        \multirow{12}*{Confusion Words} & ni hao mi & N $\text{I}_2$ H $\text{A}_3$ $\text{U}_3$ M $\text{I}_3$ \\
        & ni hao, ni hao & N $\text{I}_2$ H $\text{A}_3$ $\text{U}_3$ N $\text{I}_2$ H $\text{A}_3$ $\text{U}_3$\\
        & ni hao ya & N $\text{I}_2$ H $\text{A}_3$ YH $\text{I}_4$ $\text{A}_4$\\
        & hao mi ya & H $\text{A}_3$ $\text{U}_3$ M $\text{I}_3$ YH $\text{I}_4$ $\text{A}_4$\\
        & ni mi ya & N $\text{I}_2$ M $\text{I}_3$ YH $\text{I}_4$ $\text{A}_4$\\
        & ni hao & N $\text{I}_2$ H $\text{A}_3$ $\text{U}_3$\\
        & mi ya, mi ya & M $\text{I}_3$ YH $\text{I}_4$ $\text{A}_4$ M $\text{I}_3$ YH $\text{I}_4$ $\text{A}_4$ \\
        & hao mi, hao mi &  H $\text{A}_3$ $\text{U}_3$ M $\text{I}_3$ H $\text{A}_3$ $\text{U}_3$ M $\text{I}_3$\\
        & ni hao mi & N $\text{I}_2$ H $\text{A}_3$ $\text{U}_3$ M $\text{I}_1$\\
        & hao mi ya & H $\text{A}_3$ $\text{U}_3$ M $\text{I}_1$ YH $\text{I}_4$ $\text{A}_4$\\
        & mi ya, mi ya & M $\text{I}_1$ YH $\text{I}_4$ $\text{A}_4$ M $\text{I}_1$ YH $\text{I}_4$ $\text{A}_4$\\
        & hao mi, hao mi & H $\text{A}_3$ $\text{U}_3$ M $\text{I}_1$ H $\text{A}_3$ $\text{U}_3$ M $\text{I}_1$\\
        \bottomrule
    \end{tabular}
    }
    \vspace*{-0.4cm}
\end{table}


\section{MODEL ARCHITECTURE}
\label{sec:format}

\begin{figure}[th]
    \centering
    \includegraphics[width=0.4\textwidth]{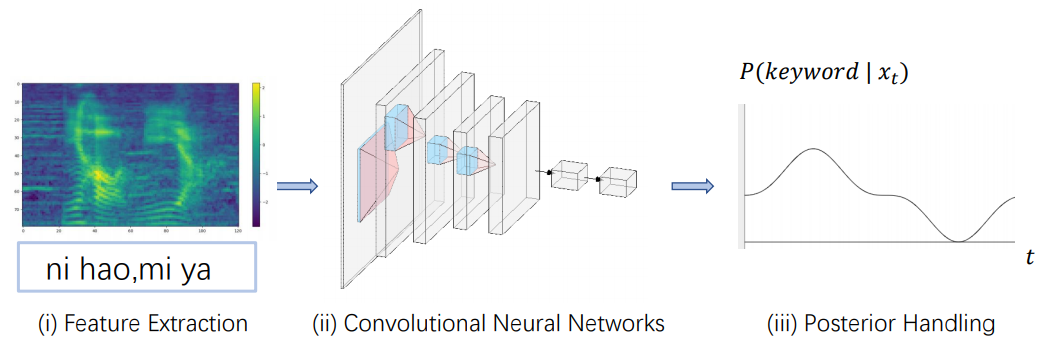}
    \caption{Framework of the baseline system.}
    \label{CNN_framework}
\end{figure}

In this section, we present our baseline system, which is modified from the CNN-based KWS system \cite{Sainath2015ConvolutionalNN}. As shown in Figure \ref{CNN_framework}, our baseline system consists of three modules:(i) a feature extraction module, (ii) a convolutional neural network and (iii) a posterior processing module. 

The feature extraction module converts the audio signals into acoustic features. 80 dimensional log-mel filterbank features are extracted from a speech frame with 50ms long and 12.5ms shift. Then we apply a segmental window with 121 frames to generate training samples that contain enough context information as the input of the model. 

Our backbone network consists of three convolutional layers each followed by a maximum pooling layer.  For all three CNN layers, the kernel size is set to (3,3), the stride is (1,1), and the pooling size is set to (2,2). Two fully connected layers and a final softmax activation layer are applied as the back-end prediction module to obtain the keyword occurrence probability. 

The acoustic feature sequence is transformed into a posterior probability sequence of selected keywords by the model. We perform the keyword detection algorithm over a sliding window with length $T_s$. Here we use $\textbf{x}^{(i)} = \{x_i, x_{i+1}, \dots , x_{i+T_s} \}$ to denote one input window over the segment $X$ that contains $N$ frames. Then the keyword confidence score is calculated as follows:
\begin{equation}
    conf(X) = \max_{1\leq t\leq N-T_s} P_{keyword}(\textbf{x}^{(t)})
\end{equation}
where $P_{keyword}(x^{(t)})$ is the posterior probability of the keyword appearing in the window started at frame $t$. The KWS system triggers once when the confidence score exceeds a predefined threshold.

\section{Adversarial Samples}
\label{sec:pagestyle}

Models that perform well on the test data set might fail in real life applications where many testing samples are confusion words. This problem becomes more important in the case of customized wake up words defined by the users. In this case, to reduce the performance degradation when applying KWS in unmatched scenarios and improve the robustness of KWS, we propose three methods to generate adversarial samples for confusion words. 

\subsection{Waveform Concatenation}

To obtain training samples of confused words, it is natural to use unit selection and waveform concatenation.\cite{concatenate} shows the difference between concatenative and neural TTS system. We use a Large Vocabulary Conversational Speech Recognition(LVCSR) to align the audio and the text in labeled public speech dataset, then truncate the audio to get waveform of each subword of the keyword. Truncated audio may come from different speakers. We simply concatenate the waveform according the order of the subwords in keywords and confused words to generate the adversarial samples. 

\subsection{Text-to-speech Augmentation}
 We obtain synthesized data from a mandarin multi-speaker TTS system \cite{Cai2020}. In this setup, 7k speakers from publicly available datasets and internal datasets are collected and used for synthesized. For each speaker, we first extract the speaker embedding with one utterance by using the TTS system. Then 3 kinds of synthesized samples are generated by the multi-speaker TTS system conditioned on the speaker embedding: (i) positive samples that whose content is the keywords (ii) negative samples that do not contains keywords, (iii) adversarial negative samples that are confusion words that have contents close to the keywords. 


\subsection{Masked Audio}

We applied random masking on keyword samples and used them as the adversarial negative data in training to improve the robustness of our KWS model. The KWS model should yield undetected results when having these masked samples since masked samples simulate confusion words like keyword audio that are interrupted by mute in the middle. For each positive sample, we generate corresponding masked samples online by replacing 40\%-60\% audio signals with Gaussian white noise, unlike SpecAug\cite{specaugment} which uses the mean value. 

\begin{figure}[t]
    \centering
    \includegraphics[width=0.45\textwidth]{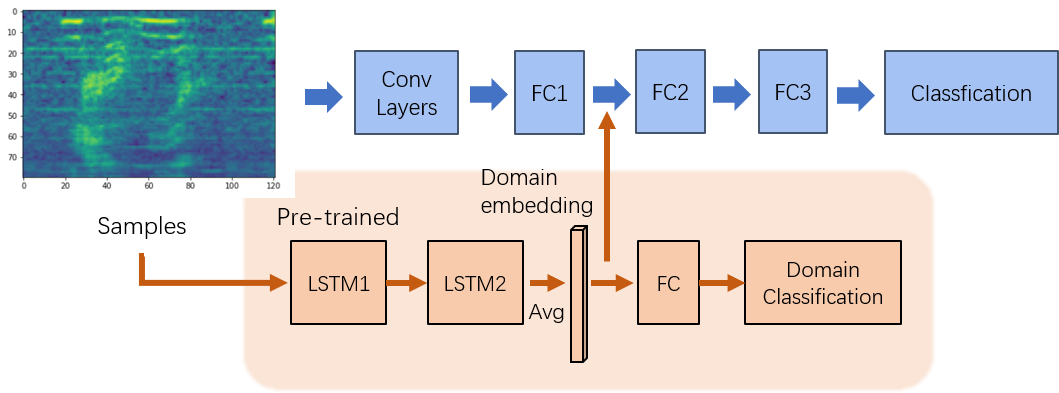}
    \caption{Framework of the domain embedding system.}
    \label{lstmconcat_figure}
\end{figure}

\subsection{Domain Adaptation}

Based on the assumption that the distribution of synthetic data and real data are different, we incorporate the domain adaptation method of Environmental Domain Embeddings with TTS augmentation in the training step in order to improve the robustness of the model and make it fit the distribution of real data better. As shown in Figure \ref{lstmconcat_figure}, we train the KWS system with the domain embedding derived from a pre-trained domain classifier. The method is inspired by \cite{embASR} and \cite{2005.03633} which applied the domain embedding to incorporate domain knowledge into network training and improved the performance of the keyword classifier on far-field conditions.

The domain classifier is trained by samples from different domain which include real domain, synthetic domain and concatenated domain. The domain classifier consists of two stacked LSTM layers followed by an average pooling layer and a final fully-connected linear layer. Domain embeddings are extracted from the output of the pooling layer and its dimension is fixed to 128. The domain classifier is trained before the training of CNNs. When we train the CNN model, we extract the domain embedding from the pre-trained domain classifier and concatenate the embedding to the output of the first fully-connected layer. Then the concatenated features are fed into two linear layer for predicting the posterior probability of the keyword.

\section{EXPERIMENTAL RESULTS}
\label{sec:typestyle}

\subsection{Dataset}
Natural speech recorded by native speakers and generated adversarial samples are both used for training in our experiments. For natural speech data, the HI-MIA dataset \cite {9054423} is used as the positive samples. The HI-MIA dataset includes speech data recorded by one close-talking microphone and six 16-channel circular microphone arrays. Each utterance contains content with four Chinese characters ``ni hao, mi ya" (Hello, Mia).  We only use the recordings from the single-channel close-talking microphone. Samples from 300 randomly selected speakers are used as the training set, and samples from 30 speakers are used as the Hi-mia test set. The Aishell-1 \cite {8384449} dataset is used as the negative sample of real speech data. Utterances from 300 speakers are selected for training, and utterances from 30 speakers are used as the Aishell-test set. 

For concatenated data, Each subword in keyword contains about 3k samples, and we concatenate the waveform online to synthesize keywords (concat-wake) and confusion words(concat-cw) as train set. 
For synthetic data, we have 7k different utterance samples from all speakers that is synthesized according to the keyword text. They are used as synthetic positive keywords(synt-wake) train set. And we also have samples that includes 12 confusion word patterns with 90k different voices, where utterance from all speakers are used as the synthetic confusion words (synt-cw) train set. Also we mask the postive samples online as negative samples according to the Section 4(mask).  
In addition, 188k negative sample audio are synthesized with provided text from  Aishell-2 \cite{1808.10583} (synt-neg). In order to compare the performance of models in practical applications, we also used the HI-MIA-CW database as the test set(real-cw). 
The statistics of the data we used for training and testing is shown in table \ref{dataset}, where the term `Real' denotes natural speech, including utterances from Hi-MIA database (Real Positive), Aishell-1 (Real Negative) and HI-MIA-CW (Real Confusion Words).

\begin{table}[h]
    \centering
    \caption{Dataset statistics
    (P: positive, N: negative)}
    \label{dataset}
    \begin{tabular}{cccccc}
        \toprule
        Samples & Label & Train & Test \\
        \midrule
        HI-MIA & P & 23k    & 2k \\
        Aishell-1 & N & 105k    & 10k \\
        HI-MIA-CW & N & - & 16k \\
        Concatenated Keywords & P & 23k & - \\
        Concatenated Confusion Words & N & 23k & - \\
        Synthesized Keywords & P & 7k & - \\
        Synthesized Confusion Words & N & 90k    & -  \\
        Synthetic Negative & N & 188k  & - \\
        Masked  & N  &   23k  & - \\
        \bottomrule
    \end{tabular}
    \vspace*{-0.4cm}
\end{table}

\subsection{Experimental Setup}
We preprocess the Hi-mia training set by trimming the beginning silence and force align the audio by a speech recognition system trained on the AISHELL-2 dataset. 
For each sample, we obtain the start time of pronouncing the word "ni" and use the following 121 frames as the final input, where 121 frames are enough for speaking the keyword 
according to the alignment information. Our models are trained for 100 epochs with Nesterov momentum Stochastic gradient descent optimizer. The initial learning rate of the optimizer is set to 0.1 and decays when the training loss has not decreased for several rounds. During evaluation, we have a sliding window with a frame length of 121 for each utterance and detect the occurrence of the expected keyword.

Six KWS systems are trained and evaluated regrading different training setups in our experiments: (i) \textbf{baseline}: use all real samples (include Real Positive set and the Real Negative set), which are shown in table \ref{dataset}, for training. (ii) \textbf{real+concat-*}:  use the all real samples, all concatenated samples(include concat-wake and concat-cw) for training. (iii) \textbf{real+syn-*}:  use all real samples, all synthetic samples(include synt-wake, synt-cw and synt-neg) for training. (iv) \textbf{real+mask}:  use all real samples, and the masked samples for training. (v) \textbf{real+concat-*+syn-*+mask}:  use all real samples, all concatenated samples, all synthetic samples, and the masked samples for training. 
(vi) \textbf{real+concat-*+syn-*+mask+EMB}:  use all real samples, all concatenated samples, all synthetic samples, and the masked samples for training. Pre-trained Domain Classifier is incorporated in this setup.



There are two combination sets for evaluation: (i) \textbf{real}: use the test set from Real Positive set and Real Negative set for evaluation. (ii) \textbf{real + real-cw}: in addition to the test sets mentioned above, the natural samples of confusion words (real-CW) are also included.

\subsection{Results}


\begin{figure}[h]
    \centering
    \includegraphics[width=0.35\textwidth]{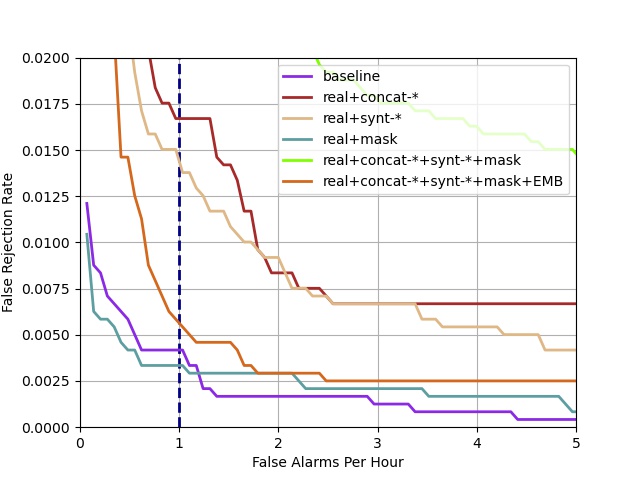}
    \caption{Performances of models on the real test sets}
    \label{fig:real_test}
\end{figure}

\begin{figure}[h]
    \centering
    \includegraphics[width=0.35\textwidth]{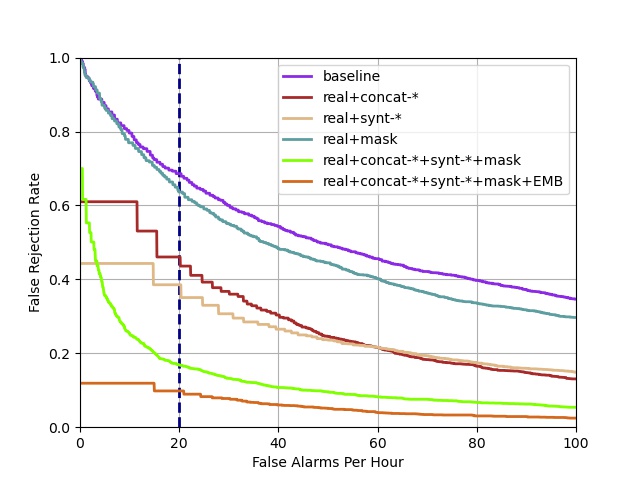}
    \caption{Performances of models on the real+real-cw test sets}
    \label{fig:realcw_test}
\end{figure}

Results are shown in Figure \ref{fig:real_test}, \ref{fig:realcw_test} and Table \ref{result_baseline1}, \ref{result_baseline2}, where Figure \ref{fig:real_test} and Table \ref{result_baseline1} shows the performance of models on real test sets (Hi-mia + Aishell1) without confusing words testing samples, Figure \ref{fig:realcw_test} and Table \ref{result_baseline2} shows performance of models on the real + real-cw test set. As for real test sets, we choose the false rejection rate under one false alarm per hour as each model's performance criterion separately. Table \ref{result_baseline2} presents the KWS performance of five models regarding the false rejection rate when the false alarm rate per hour is 20, as adding confusion words in the test set makes the task more challenging.

\begin{table}[h]
    \centering
    \caption{Performances of models trained with different methods on the real test sets (the false rejection (FR) rate (\%) under one false alarm (FA) per hour)}        
    \label{result_baseline1}
    \begin{tabular}{ccccc}
        \toprule
        Training set & real  \\
        \midrule
        baseline    & 0.417   \\
        \midrule
        real + concat-* & 1.67   \\
        real + syn-*  & 1.37    \\
        real + mask    & \textbf{0.334}  \\
        real + concat-* + synt-* + mask  & 3.29     \\
        real + concat-* + synt-* + mask + EMB & 0.523     \\
        \bottomrule
    \end{tabular}
    \vspace*{-0.4cm}
\end{table}

\begin{table}[h]
    \centering
    \caption{Performances of models trained with different methods on the real+real-cw test sets (the false rejection (FR) rate (\%) under twenty false alarms per hour)}        
    \label{result_baseline2}
    \begin{tabular}{ccccc}
        \toprule
        Training set & real+real-cw   \\
        \midrule
        baseline    & 68.60   \\
        \midrule
        real + concat-* & 46.05 \\
        real + syn-*  & 38.54    \\
        real + mask    & 63.63  \\
        real + concat-* + synt-* + mask  & 16.87     \\
        real + concat-* + synt-* + mask + EMB & \textbf{9.81}     \\
        \bottomrule
    \end{tabular}
    \vspace*{-0.4cm}
\end{table}

From Table \ref{result_baseline1} and \ref{result_baseline2} we can obtain the following observations. First, the baseline system performed well in real test sets without confusing word samples. However, the baseline system's performance will degrade dramatically on confusion words examples, which is frequently happened in real-life applications. Second, directly adding adversarial samples will lead to performance degradation on the real test set but masked samples help train the system and achieve best result 0.334 on the real test set. Moreover, after the domain embedding algorithm is applied, the system also maintain the performance on the real test set and achieve the result 0.523. It is because domain adaptation methods help the system to learn to fit the distribution of real data better and overcome the the degradation in performance due to the domain shift.
Third, the confusion word test set's accuracy has been significantly improved by adding adversarial synthetic samples. It can also be found that adding concatenated samples and adding masked samples does not improve the performance as much as TTS synthesized ones. 
This may be due to insufficient simulation of confusion word when masked samples are added separately, while possibly misleading the system to learn whether there is a Gaussian distribution of judgments. Also the concatenated samples show steep changes in the splicing breakpoints in the spectral features and do not simulate the confusion words well enough. 
Fourth, adding synthetic samples along with concatenated and masked samples can help the system to better learn the difference between confusing words and keywords, which achieve the result 16.87\% on the real+real-cw testing set.

Finally, comparing to the baseline without any augmentation, this augmentation setup with the domain adaptation method achieves best performance on the real+real-cw testing set and shown great robustness on confusion words scenarios as the false rejection rate under twenty false alarms per hour decreases from 68.60\% to 9.81\%. 



\section{CONCLUSIONS}
\label{sec:majhead}

In this paper, we discuss the concept of the confusion words and focus on the task of small-footprint keyword spotting in this scenarios, then show the effectiveness of generating the adversarial samples to train a keyword recognition system. 
Confusing words that sound very similar to the keywords lead to a significant degradation in system performance. We release the supplemental database HI-MIA-CW and adopt three augmentation strategies to enhance the robustness, including concatenated samples, masked samples and synthesized samples with the domain adaptation methods. Experimental results show that our proposed methods can effectively maintain the accuracy on general real test data and at the same time, achieve significant improvement under the test condition with confusing word samples.





\section{ACKNOWLEDGMENTS}
\label{sec:copyright}

This research is funded in part by the National Natural Science Foundation of China (62171207), the Fundamental Research Funds for the Central Universities (2042021kf0039), Key Research and Development Program of Jiangsu Province (BE2019054), Science and Technology Program of Guangzhou City (201903010040, 202007030011) and Lenovo.




\bibliographystyle{IEEEbib}
\bibliography{strings,refs}

\end{document}